\documentclass{aa}  

\usepackage{graphicx}
\usepackage{txfonts}
\usepackage{multirow}
\usepackage{natbib}
\newcommand{\masyr}{${\rm mas\, yr^{-1}}$}

\newcommand{\gaia}{{\it Gaia~}}
\begin{document}

   \title{With and without spectroscopy: {\it Gaia} DR2 proper motions of seven Ultra-Faint Dwarf Galaxies}

   \subtitle{}

   \author{D. Massari
          \and
          A. Helmi
          }

   \institute{Kapteyn Astronomical Institute, University of Groningen, NL-9747 AD Groningen, Netherlands\\
              \email{massari@astro.rug.nl}     
             }

   \date{Received 4/5/2018; accepted 15/10/2018}

 
  \abstract
   {}
   {We present mean absolute proper motion measurements for seven ultra-faint dwarf galaxies orbiting the Milky Way, namely Bo\"{o}tes~III, 
   Carina~II, Grus~II, Reticulum~II, Sagittarius~II, Segue~2 and Tucana~IV. For four of these dwarfs our proper motion estimate is the first
   ever provided.}
 {The adopted astrometric data come from the second data release of
   the {\it Gaia} mission. We determine the mean proper motion for
   each galaxy starting from an initial guess of likely members, based
   either on radial velocity measurements or using stars on the
   Horizontal Branch identified in the {\it Gaia} ($G_{\rm BP}$-$G_{\rm RP}$,
   G) colour-magnitude diagram in the field of view towards the UFD.  We then refine their membership 
   iteratively using both astrometry and photometry.  We take into
   account the full covariance matrix among the astrometric parameters
   when deriving the mean proper motions for these systems.}
 {Our procedure provides mean proper motions with typical
   uncertainties of $\sim0.1$~\masyr, even for galaxies without prior
   spectroscopic information. In the case of Segue 2 we find that
   using radial velocity members only leads to biased results,
   presumably because of the small number of stars with measured
   radial velocities.}
   {Our procedure allows to maximize the number of member stars per galaxy regardless of the existence of prior spectroscopic information, and can therefore
   be applied on any faint or distant stellar system within reach of {\it Gaia.}}
  
   \keywords{Astrometry – Proper motions –  Galaxies: dwarf – Galaxies: Local Group – galaxy: kinematics and dynamics}

   \maketitle
%

\section{Introduction}

The orbits of the satellite galaxies of the Milky Way constitute a
powerful tool to answer several important questions of modern
astrophysics. They provide valuable constraints on the mass and shape
of the Milky Way halo (\citealt{little87, we99}), as well as on the
evolution of the satellites themselves and their resilience to tidal
forces (\citealt{pena08, lokas12}). Orbits also 
allow us to investigate whether their dynamical evolution and star
formation histories are connected (\citealt{grebel01, tolstoy04,
  battaglia08}). Furthermore, they reveal how our Galaxy has
assembled its population of dwarf galaxy satellites
(\citealt{kroupa05, li08}), and how this process relates to the large
scale environment in which the Galaxy is embedded (\citealt{libeskind05, buck16}).

To determine the orbits of Milky Way satellites, their position on the
sky, distance, line of sight velocity and proper motions are required.
Historically, the last two observables of this six-dimensional phase
space have been the most difficult to measure. Only recently and
mostly thanks to the exquisite astrometric capabilities of the {\it
  Hubble Space Telescope} (HST), absolute proper motions have been
provided for many of the brightest dwarf galaxies orbiting our Galaxy
(e.g. \citealt{piatek03, massari13, kalli13, sohn13, sohn17}).

The advent of the \gaia mission (\citealt{prusti16, brown16}) has
resulted in a quantum leap in our ability to measure proper motions
for distant stellar systems.  Already with its first data release,
thanks to the combination with pre-existing datasets such as {\sc
  TYCHO}, HST or the Sloan Digital Sky Survey, \gaia enabled
measurements of the proper motions of Galactic globular clusters
(\citealt{massari17, watkins17}), stellar streams (\citealt{helmi17,
  deason18}) and stars in dwarf spheroidal galaxies
(\citealt{massari18}).  Yet, it is with the second data release
(\citealt{brown18}), that \gaia becomes
transformational. The \gaia DR2 catalogue contains absolute proper
motions for more than one billion stars, and this has permitted for
the first time to directly measure proper motions over the full sky
and without the need of external objects for absolute calibration.
This has resulted in spectacularly precise proper motion estimates for
75 Galactic globular clusters, the Large and the Small Magellanic
Clouds, the nine classical dwarf spheroidal galaxies, and one
ultra-faint dwarf galaxy \citep{helmi18}.

Ultra-faint galaxies (UFDs) are very difficult objects to detect because of their low
luminosity, but the number of known UFDs is continuously increasing
thanks to wide field deep surveys (e.g. \citealt{dw15, laevens15,
  torrealba18}). UFDs might well constitute one of the best solutions
to the missing-satellites problem (\citealt{moore99, klypin99}). 
Despite their cosmological importance, very little is
known about their origin, and knowledge of their orbits around the
Milky Way is a powerful way to find clues. Until very recently, only
two UFDs had an absolute proper motion measurement (\citealt{fritz17,
  helmi18}). Around the time of submission of this Paper, several new
investigations reported absolute proper motions for UFDs using
stars defined as members based on radial velocity measurements
(\citealt{simon18, fritz18, kalli18, carlin18}).  However, relying on
spectroscopic information only is not entirely advisable. It limits
the number of UFDs whose proper motions can be determined, as not all
UFDs have been followed up spectroscopically (see also \citealt{pace18}). 
On the other hand, as we show here, it might lead to biased results 
because of small numbers statistics.

In this paper we present the absolute proper motion for seven
UFDs, namely Bo\"{o}tes~III, Carina~II, Grus~II, Reticulum~II, 
Sagittarius~II, Segue~2 and Tucana~IV. All of these are located within 70 kpc from us and have an
integrated absolute V-band magnitude brighter than M$_{\rm V}=-2.5$. Four out
of this seven UFDs do not have publicly available spectroscopic
information. We develop a method to reliably select likely members
even in such cases, and show that this method also fares
better on UFDs with spectroscopic information, because it allows to
infer a more complete sample of members that is not limited to stars with
measured radial velocities.

\section{Stellar membership and mean proper motion determination}\label{method}

In a recent paper, \cite{simon18} determined the mean proper motion
for 17 Ultra-faint dwarfs (UFDs) by using all the stars defined as
members according to their radial velocities. However, this kind of
information is not available for the entire sample of Milky Way dwarf
satellites. Moreover, spectroscopic samples are incomplete by nature
being limited to small numbers of stars, and this can potentially lead
to biased mean proper motion estimates (see below). For these reasons,
in this paper we develop a procedure that is less prone to these
issues, and which we use to determine stellar membership and mean
proper motions of each of the analysed UFDs. 

Our procedure is iterative and consists of the following steps that are applied to each one of the UFDs:
\begin{enumerate}
\item We identify initial likely members from either spectroscopic
  (radial velocity member candidates) or photometric (Horizontal
  Branch candidates) information.
\item From the initial set of likely members, we estimate mean
  proper motions ($\langle\mu_{\alpha}\cos\delta\rangle$, $\langle\mu_\delta\rangle$) and
  parallax $\langle\varpi\rangle$, together with the corresponding standard
  deviations ($\sigma_{\mu\alpha}$, $\sigma_{\mu\delta}$,
  $\sigma_{\varpi}$).
 \item We apply a $2.5\sigma$ selection around these mean astrometric parameters.
 \item We perform a further selection on $i$) the colour-magnitude diagram
   (see e.g. the blue box in Fig.~\ref{sel_hb}), $ii$) projected
   distance from the centre of the UFD, and $iii$) we exclude stars
   with $0<\sigma_{\varpi}/\varpi<0.2$ to remove
   obvious very nearby foreground contaminants.
 \item The stars that survive these selection criteria are defined as new likely members, and their mean astrometric parameters and associated uncertainty (defined as the 
 root mean square around the mean) are determined.
 \item We repeat steps 3, 4 and 5 until convergence is reached. Convergence is defined when the change between the mean astrometric parameters
 in two subsequent iterations is smaller than $0.5$ times the previously determined root mean square (rms) error. 
\end{enumerate}

The convergence of the whole procedure, especially for the cases of
poorly populated UFDs or when the contamination of the surrounding
field is strong, depends on the adopted projected distance cut.  
  The choice of the projected distance cut results from a balance
  between avoiding excessive contamination and including the largest
  number possible of members. We find that the best solution is different 
  for each single UFD, and we applied cuts ranging from $\sim3.5$\arcmin~
  for the most crowded case (Sagittarius~II) to
  $\sim20$\arcmin~ for the most diffuse UFDs (see the red circles in
  Fig.~\ref{sky_rv} and Fig.~\ref{sky_hb}.)  The mean proper motions
obtained in the last step are adopted as the final solution.

We underline that the mean proper motions of the UFDs have been
computed by taking into account the correlations among the astrometric
parameter uncertainties for each of the stars following the
  method developed by \cite{fvl09} and described in detail in Appendix
  A.1 of \cite{fvl17}\footnote{Note that we neglected the terms in the
    covariance matrix related to the UFDs instrinsic parallax and
    proper motion dispersion because our targets are tens of kpc away
    from us.}. We find that not properly including these
correlations, results in systematic errors that can be as large as
$\sim0.05$ \masyr.

In the next section we present the results obtained by following this 
procedure when applied on our sample of seven UFDs. We first focus on the dwarfs
that have spectroscopic follow-up such that the initial guess is based
on radial velocity determinations, and then on those UFDs for which we
use horizontal branch stars. In Table~\ref{tab1} we list the resulting
proper motions $\langle\mu_{\alpha}\cos\delta\rangle$,
$\langle\mu_\delta\rangle$, the associated uncertainties
(defined as the rms error around the mean), covariance and number of final
members used for all of the seven UFDs.  We stress that, though not reported in this analysis,
systematic errors on the mean proper motion estimates as large as
$\sim0.035$~\masyr are expected to act on the small scales sampled in
this work (see e.g. Sect.~4.1 in \citealt{helmi18}, Sect.~4.2 in \citealt{arenou18}
and Sect.~5.4 in \citealt{lindegren18}).
The list of IDs and positions of our final members for each UFD is made publicly available through the 
Centre de Données astronomiques de Strasbourg (CDS\footnote{http://cdsweb.u-strasbg.fr/cgi-bin/qcat?J/A+A/}). 
An example showing the first four members of each
galaxy is given in Table~\ref{tab2}.

\section{Results: UFDs with spectroscopic information}\label{wspec}

\begin{figure}
    \includegraphics[width=\columnwidth]{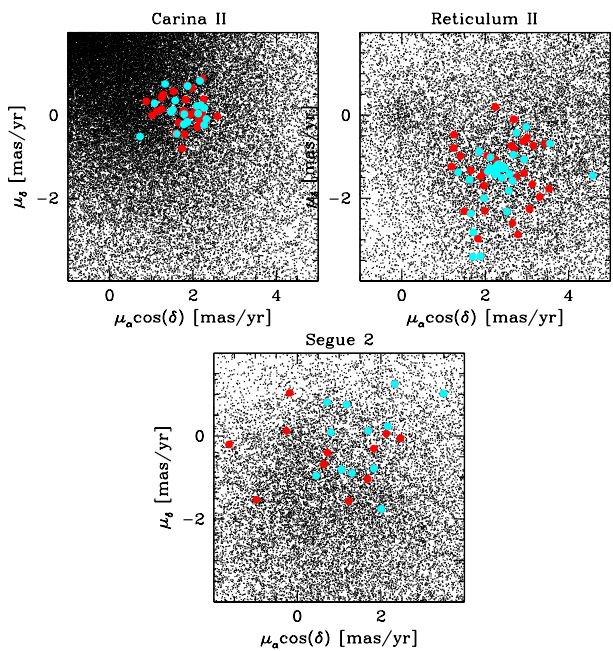}
    \caption{\small VPDs for the stars in the direction of the three
      UFDs with published radial velocity catalogues. Cyan filled
      circles mark the spectroscopic members. Red filled circles
      highlight all the stars defined as members at the end of the
      procedure described in Sect.~\ref{method}. Black dots show all
      the stars in a field of view of 3 degrees radius from the
      centres of the UFDs.}\label{vpd_rv}
\end{figure}

In our sample of dwarfs, Carina~II, Reticulum~II and Segue~2 have
radial velocity measurements available in the literature (see \citealt{li18, simon15,
  koposov15, kirby13}).  Therefore, for these three dwarfs we
cross-matched their membership lists with the \gaia DR2 catalogue and used
the stars in common for the first step of our procedure. The location
of these stars in the Vector Point Diagrams (VPDs) is shown as cyan
filled circles in Fig.~\ref{vpd_rv}.

\begin{figure}
    \includegraphics[width=\columnwidth]{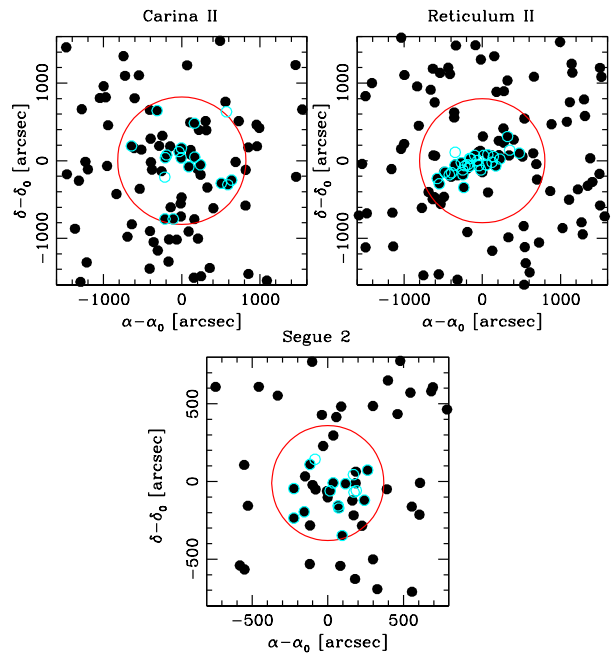}
        \caption{\small Location on the sky of stars surviving the last iterative step of the membership procedure. For sake of representation,
        the distance cut has not been applied but is marked with a red circle. Cyan empty circles mark initial radial velocity candidate members.}\label{sky_rv}
\end{figure}

We then applied our procedure which converges quickly, i.e.\ after 2-3 
iterations. The final sample of member stars are marked by the red
filled circles in the same figure.  As expected this figure shows 
that the number of members we find is larger than that coming only
from the spectroscopic data.  Moreover, a few of the radial velocity
candidate members turned out to be false members, and these are shown
with the cyan empty circles in Fig.~\ref{sky_rv}. These stars are
excluded from the final computation of the mean proper
motions, which are therefore more accurate also because of the larger
number of true members. 

The step-wise procedure we follow thus constitutes a big improvement,
and the case of Segue 2 highlights another reason why.  For this
system the distribution of spectroscopic members in the VPD is
significantly shifted towards larger $\mu_{\alpha}\cos\delta$ and
larger $\mu_{\delta}$ in comparison to the rest of the stars we
identify as members. This results in  a small discrepancy
between \cite{simon18} and our own measurement: our derived mean
proper motion is ($\langle\mu_{\alpha}\cos\delta\rangle$,
$\langle\mu_\delta\rangle$) $=$ ($1.27\pm0.11$, $-0.10\pm0.15$)
\masyr, which is $\sim 1.5\sigma$ away from the value quoted in
\cite{simon18}.

On the other hand, the VPD distribution of spectroscopic members for
Carina~II and Reticulum~II seems to represent well the overall
population as determined by our procedure.  In fact, our results match
within $1\sigma$ those in \cite{simon18, fritz18, kalli18}, being
($\langle\mu_{\alpha}\cos\delta\rangle$, $\langle\mu_\delta\rangle$)
$=$ ($1.81\pm0.08$, $0.14\pm0.08$) \masyr for Carina~II and
($\langle\mu_{\alpha}\cos\delta\rangle$, $\langle\mu_\delta\rangle$)
$=$ ($2.34\pm0.12$, $-1.31\pm0.13$) \masyr for Reticulum~II.  The
  small remaining differences might be explained by differences in the
  (often larger) sample of members produced by our procedure, and 
  by the fact that we took into account the correlations among the
  astrometric parameters (see \citealt{arenou18} and Fig.~A.11 in
  \citealt{helmi18}), as well as by the use of a standard mean instead of a
  weighted mean.

The distribution on the sky of our selected members is a consistency
check validating the robustness of our selection procedure, and also
extremely interesting from a scientific point of view. We should
reasonably expect that, if our selection works properly, the selected
members of each galaxy define an overdensity with respect to field
stars in their distribution in Right Ascention ($\alpha$) and Declination
($\delta$). This is indeed nicely shown in Fig.~\ref{sky_rv}.

For all the three UFDs, the members (black filled circles within the
projected distance cut shown as a red circle) define quite elongated
or asymmetric structures, suggestive of tidal effects acting on these
UFDs and affecting their shape. Such features were already recognised
in the Magellanic Satellite Survey photometry of Carina~II
(\citealt{torrealba18}), the Dark Energy Survey photometry of
Reticulum~II (\citealt{bechtol15}) and the Sloan Extension for
Galactic Understanding and Exploration of Segue~2
(\citealt{belokurov09}). While in those cases the field
decontamination was based on statistical methods, here it relies on
kinematical observed properties and it is therefore more
robust. Nonetheless the similarities between the shapes of the sky
distributions obtained in the different ways are remarkable.

\section{Results: UFDs without spectroscopic information}

\begin{figure}
    \includegraphics[width=\columnwidth]{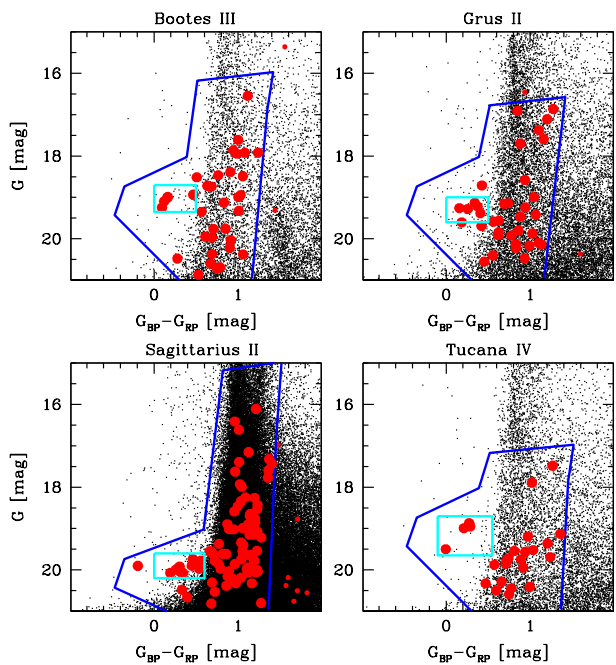}
        \caption{\small CMDs for the field of view around the four UFDs with no available spectroscopic information. The cyan box highlights the region where 
        candidates HB have been initially selected. Red filled circles mark the final list of members, again without applying the CMD selection but representing 
        it with the blue box.}\label{sel_hb}
\end{figure}

The UFDs Bo\"{o}tes~III, Grus~II, Sagittarius~II and Tucana~IV do not
have publicly available spectroscopic information. As mentioned in
Sect.~\ref{method} we therefore use Horizontal Branch (HB) candidates
located within $15$\arcmin~ from the centre of each UFD (from \citealt{mc12}) to
obtain an initial guess on the mean astrometric parameters.  HB
stars have been chosen because they are easy to detect in the CMD,
being bluer than the typical sequences described by the majority of
stars located in the respective fields of view. This is clearly
demonstrated in Fig.~\ref{sel_hb}, where the CMD regions adopted to
select HB stars are marked with a cyan box.

To demonstrate the effectiveness of the selection procedure described
in Sect.~\ref{method}, Fig.~\ref{sel_hb} shows a snapshot of the last
iteration step from the point of view of the CMD. The red filled
circles are stars that survived the $2.5\sigma$-clipping astrometric
selection and items 4$ii$) and 4$iii$) of the procedure.  However,
only those located inside the blue delimited box will be used for the
final mean proper motion estimate.  In general, the selected members
are located on the typical evolutionary sequences characteristic of the
stellar populations seen in dwarf galaxies. These sequences are quite well-defined despite
the meagre numbers ($40$-$50$ stars). The fact that the poorly populated red giant branches are
also quite faint (the brightest HB is at $G \sim19$) contributes in
spreading them out in $G_{\rm BP}-G_{\rm RP}$ colour. The colour uncertainty
at these magnitudes is $\sim 0.1$, whereas
the measured spread amounts to $\sim 0.15$. Therefore we cannot 
exclude the presence of multiple, possibly chemically heterogeneous
sub-populations (in fact UFDs
typically show spread in their metallicity distribution, see
\citealt{tolstoy09}). 

\begin{figure}
    \includegraphics[width=\columnwidth]{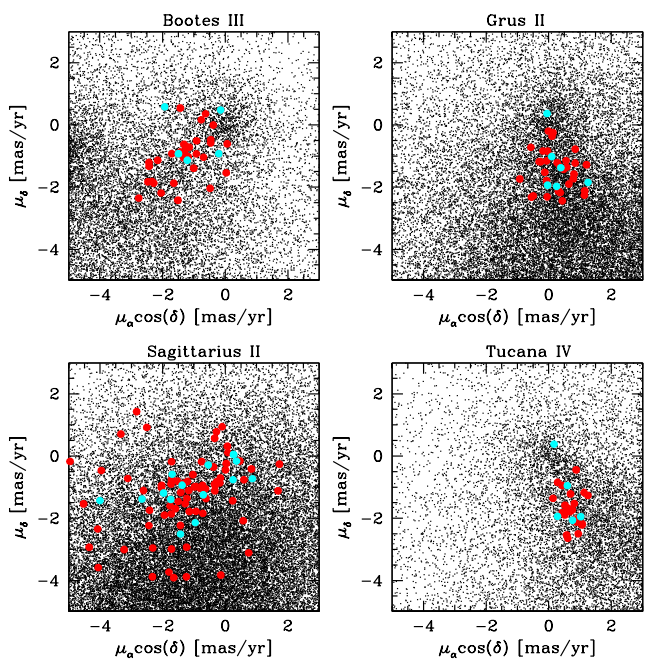}
        \caption{\small VPDs for the stars around the four UFDs. Red filled circles mark stars belonging to the final list of members.}\label{vpd_hb}
\end{figure}

The distribution in the VPD of the final members of the four UFD
galaxies is shown in Fig.~\ref{vpd_hb} with red filled circles.  Like
for the dwarfs analysed in Sec.~\ref{wspec}, they describe a
reasonably well concentrated clump of stars, clearly separated from
the bulk of the field population (black dots). It is interesting to
note that in the case of Bo\"{o}tes~III, a third concentration of
stars clearly stands out at $(\langle\mu_{\alpha}\cos\delta\rangle$,
$\langle\mu_\delta\rangle) \sim (-5, -1)$~\masyr: it is made up of
stars belonging to the Galactic globular cluster NGC~5466, located
very close to the galaxy also on the sky. Their quite different
distances from the Sun (with the UFD $~\sim 30$ kpc farther away,
\citealt{ferraro99}) and kinematics, with Bo\"{o}tes~III having
($\langle\mu_{\alpha}\cos\delta\rangle$, $\langle\mu_\delta\rangle$)
$=$ ($-1.21\pm0.13$, $-0.92\pm0.17$)\masyr, corresponding to a
difference in velocity of about $100$ km~s$^{-1}$, implies that the
two systems are not associated to each other.

The spreads in the distributions of members in the VPDs of the four
UFDs are consistent with the typical proper motion uncertainties of
their stars. This is true even for the case of Sagittarius~II, whose distribution
seems significantly more extended and asymmetric than for the other systems. This is because 
for its stars the errors on $\mu_{\alpha}\cos\delta$ are 
a factor of $\sim1.5$ larger than those on $\mu_{\delta}$. The cyan filled
circles in Fig.~\ref{vpd_hb} mark the initial HB candidates, most of them being located
well within the region of the final members.  

\begin{figure}
    \includegraphics[width=\columnwidth]{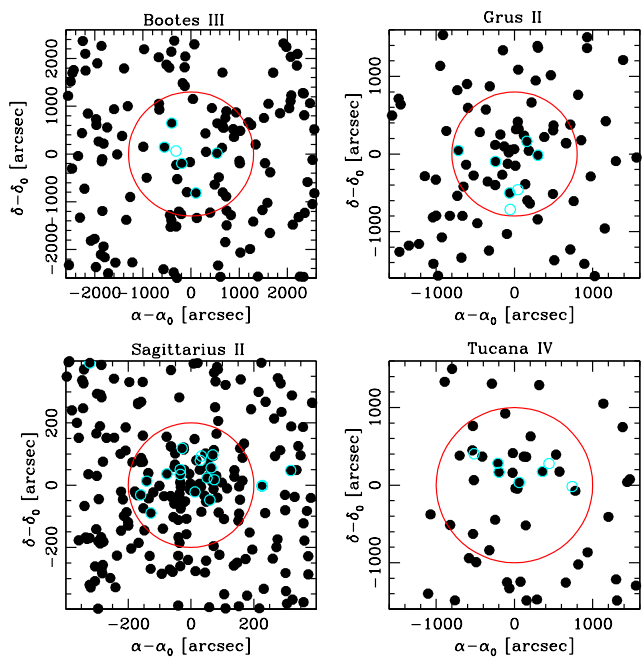}
        \caption{\small Same as in Fig.~\ref{sky_rv} for the four UFDs without spectroscopic information.}\label{sky_hb}
\end{figure}

Finally, Fig.~\ref{sky_hb} shows the distribution of the selected
members on the sky. As in Sect.~\ref{wspec}, only stars represented as
black dots and within the red circle have been considered for the
final mean proper motion estimate. All the four galaxies show a peak
in the concentration of their members within the projected distance
cut given by the radius of the red circle.  Interestingly however, for
some of these (namely Bo\"{o}tes III and Tucana IV), the peak of the
density distribution seems somewhat offset with respect to the nominal
centre (given by the centre of the red circle, and taken from
\citealt{grillmair09} and \citealt{dw15}, respectively). Moreover all
the systems, with maybe the exception of Sagittarius II (which has
been proposed to be a globular cluster rather than a UFD, see
\citealt{mp18}), appear asymmetric and possibly depict an
elongated shape.  It is beyond the aim of this paper to make a
quantitative analysis on these aspects, yet our findings already
highlight how powerful \gaia can be to better assess the density
profiles and centres of systems like the UFDs, which suffer from large
amounts of contamination by field stars.

The location of all the initially selected candidates HB stars are
marked in Fig.~\ref{sky_hb} with cyan empty circles. Few of these HB
stars have been lost during the procedure, but this was to be expected
given that the initial selection is astrometrically blind. The fact
that the actual members all lie well within the adopted projected
distance cut from the nominal centre is a further check on the
robustness of our procedure. Sagittarius~II might show hints of the
presence of few HB members outside this distance cut. Increasing such
a radius does not improve the final proper motion estimate, and given
the high degree of contamination in that field we cannot conclude they
are false members.

\section{Summary}\label{summary}

In this paper we have estimated the mean absolute proper motion for seven UFD satellites of the Milky Way, 
all located within 70 kpc from us and with an integrated V-band magnitude brighter than V=15.
This has ensured that a sufficiently large number of members are above \gaia magnitude limit of  $G\simeq21$.
For four of these seven dwarfs, ours is the first proper motion estimate ever provided.

To determine likely member stars from which to compute the
mean proper motion of each system we developed a new procedure 
which proved to be less prone to systematic uncertainties than using only 
radial velocity members in case of very low number statistics. 
This procedure starts from an initial
guess on the mean proper motion and parallax coming from an initial
sample of tentative members, and then refines their membership
and the mean astrometric parameters by coupling photometric
(CMD selection) and astrometric information together.  The initial
sample of candidate members is made up of spectroscopic members,
if radial velocity measurements are publicly available, or HB stars,
otherwise.

For all the seven UFDs studied here, our procedure works well. We
obtain well-defined evolutionary sequences in the CMDs and a clear
concentration of stars in the VPDs. Interestingly, the distribution on
the sky of the selected members often highlights elongated or
asymmetric structures, suggestive of tidal perturbations or complex
dynamical histories, and a couple of galaxies might be offset with
respect to their previously determined density centres. Future
investigations based on \gaia photometric and astrometric datasets,
possibly in combination with dynamical models, will be used to assess
these findings.

When comparing our results to those of \cite{simon18} and
\cite{fritz18} for the sub-sample of UFDs with spectroscopic information, 
we find good agreement for Carina~II and Reticulum-II, where small differences can
be ascribed to the different sample of members used (ours not being
limited to spectroscopic members only) and to the fact that we took
into account the full covariance matrix in the determination of the
mean proper motions.  On the other hand, we found a $\sim1.5\sigma$
difference with the results of \cite{simon18} for Segue 2, a system
that this author recognised as peculiar due to the wide spread of its
stars in the VPD. In this case, the sub-sample of spectroscopic
members turns out to be systematically shifted in the VPD. This
demonstrates that when limited to small samples systematic effects
related to small numbers statistics, might bias the results.

  In the case of UFDs without public spectroscopic information, we
  find good agreement for Bo\"{o}tes~III with the results of
  \cite{carlin18}, who used proprietary radial velocities to assess the
  stellar membership and found a mean proper motion which matches ours
  well within a 1$\sigma$ uncertainty. This is possibly the best
  confirmation on the validity of our method.  Finally, our estimates
  on Grus~II and Tucana~IV agree well with those of \cite{pace18},
  both in terms of mean proper motions and of expected number of
  members.

We have shown that thanks to {\it Gaia}, measurements of the absolute
proper motions of many of the known UFD satellites of the Milky Way
are within reach. This means that several of the questions relating to
the nature and evolution of these systems will likely soon be
answered.
  
\begin{table*}
\centering
\caption{Results for the seven UFDs analysed in this paper. We remark that an additional systematic error of $0.035$ mas/yr should be added to the quoted
uncertainties, as explained in \cite{helmi18}.}\label{tab1}
\begin{tabular}{lccccccc}
\hline
Name&Initial &$\langle\mu_{\alpha}\cos\delta\rangle$&$\sigma_{\mu\alpha}$&$\langle\mu_{\delta}\rangle$&$\sigma_{\mu\delta}$&cov($\mu_{\alpha}\cos\delta$, $\mu_{\delta}$)&N$_{members}$\\
    &       guess      &          \masyr          &        \masyr      &     \masyr     &      \masyr  &     \\
\hline
        Carina~II       &   RV  &  1.81  & 0.08 &  0.14 & 0.08 &  0.01&  39\\
        Reticulum~II    &   RV  &  2.34  & 0.12 & -1.31 & 0.13 &  0.04&  54\\
        Segue~2         &   RV  &  1.27  & 0.11 & -0.10 & 0.15 &  0.15&  22\\
        Bo\"{o}tes~III  &   HB  & -1.21  & 0.13 & -0.92 & 0.17 &  0.23&  34\\
        Grus~II         &   HB  &  0.22  & 0.12 & -1.41 & 0.11 & -0.02&  32\\
        Sagittarius~II  &   HB  & -1.18  & 0.14 & -1.14 & 0.11 &  0.24&  78\\ 
        Tucana~IV       &   HB  &  0.72  & 0.09 & -1.71 & 0.10 &  0.00&  23\\
\hline
\end{tabular}
\end{table*}

\begin{table*}
\centering
\caption{List of IDs and sky positions for stellar members of each UFD.}\label{tab2}
\begin{tabular}{lccc}
\hline
Name & ID & $\alpha$ & $\delta$    \\
     &    & deg      &  deg        \\
\hline
        Carina~II       &     5293875465260936320 & 114.049350695341104 & -58.207932320214731 \\
        Carina~II       &     5293875602699890048 & 113.993189519658515 & -58.207668701483918 \\
        Carina~II       &     5293954286501519616 & 114.191046317612503 & -57.865022276137211 \\
        Carina~II       &     5293954252141786240 & 114.162914972205868 & -57.862210528571794 \\
        Reticulum~II    &     4732507468553979392 &  53.837424462541811 & -54.063363888485966 \\
        Reticulum~II    &     4732508327547439744 &  53.760480435347574 & -54.065062817767739 \\
        Reticulum~II    &     4732506781359189632 &  53.695115070388930 & -54.089402284881487 \\
        Reticulum~II    &     4732587324881689472 &  54.252496159348958 & -54.061444772027166 \\
        Segue~2         &       87200583572213248 &  34.883527968906876 &  20.096188684137729 \\
        Segue~2         &       87214430546419968 &  34.750180878254781 &  20.162699696363525 \\
        Segue~2         &       87215465633916672 &  34.894480474485881 &  20.195648078594573 \\
        Segue~2         &       87213713287264000 &  34.815019180246509 &  20.158030220196210 \\
        Bo\"{o}tes~III  &     1450854452398921600 & 209.521895346528595 &  27.009750666876087 \\
        Bo\"{o}tes~III  &     1450830194423500416 & 209.615140787363885 &  26.745659458321608 \\
        Bo\"{o}tes~III  &     1450828682594980736 & 209.479318442451046 &  26.664838037392862 \\
        Bo\"{o}tes~III  &     1450806009463130368 & 209.273455059705213 &  26.578098208968804 \\
        Grus~II         &     6561433598368152192 & 331.083870655932799 & -46.394873384318629 \\
        Grus~II         &     6561421778618145792 & 330.920582092078746 & -46.407599446719615 \\
        Grus~II         &     6561421645474640000 & 330.966732559516743 & -46.412970323100460 \\
        Grus~II         &     6561410199386304384 & 331.018449330896260 & -46.422660051895832 \\
        Sagittarius~II  &     6864048339690750336 & 298.120671436779446 & -22.076789624999122 \\
        Sagittarius~II  &     6864048648924739328 & 298.127874203478086 & -22.057258648397475 \\
        Sagittarius~II  &     6864048408406508416 & 298.159108858091372 & -22.069816686574047 \\
        Sagittarius~II  &     6864047579475717760 & 298.183112466938894 & -22.082904155645384 \\
        Tucana~IV       &     4905859740259086336 &   0.662980757245541 & -60.593438485522903 \\
        Tucana~IV       &     4905641242387742976 &   0.545088807160125 & -61.082761958793881 \\
        Tucana~IV       &     4905854582002922112 &   0.718188811757548 & -60.735354695502608 \\
        Tucana~IV       &     4905852073742024192 &   0.424073005459447 & -60.725479804804138 \\
\hline
\end{tabular}
\end{table*}

\begin{acknowledgements}
  We thank the referee for their report and suggestions which helped 
  us to improve the quality of the paper.
  DM and AH acknowledge financial support from a Vici grant from NWO.
  This work has made use of data from the European Space Agency (ESA)
  mission \gaia (http://www.cosmos.esa.int/gaia), processed by the
  \gaia Data Processing and Analysis Consortium (DPAC, {\tt
    http://www.cosmos.esa.int/web/gaia/dpac/consortium}). Funding for
  the DPAC has been provided by national institutions, in particular
  the institutions participating in the \gaia Multilateral Agreement.

\end{acknowledgements}

%
\bibliographystyle{aa} 

\end{document}